\documentclass[twocolumn,aps,floatfix]{revtex4-1}
\usepackage{amssymb}
\usepackage{graphicx}
\usepackage{amsmath}
\usepackage[T1]{fontenc}
\usepackage{pstricks}

\begin{document}
\title{Optimization as a route towards observing \\the Einstein-de Haas effect in a rubidium condensate}

\author{Tomasz \'Swis{\l}ocki,$^1$ Mariusz Gajda,$^{1,2}$ and Miros{\l}aw Brewczyk$^{2,3}$}

\affiliation{
\mbox{$^1$Institute of Physics PAN, Al. Lotnik\'ow 32/46, 02-668 Warsaw, Poland}  \\
\mbox{$^2$Center for Theoretical Physics PAN, Al. Lotnik\'ow 32/46, 02-668 Warsaw, Poland} \\
\mbox{$^3$Wydzia{\l} Fizyki, Uniwersytet w Bia{\l}ymstoku, 
                             ul. Lipowa 41, 15-424 Bia{\l}ystok, Poland}    }

\date{\today}

\begin{abstract}

The main obstacle in experimental realization of the Einstein-de Haas effect in a Bose-Einstein condensate is necessity of a very  precise control of the extremely small (of the order of tens of $\mu$G) external magnetic field. In this paper we numerically study the response of a rubidium condensate to an optimized time-dependent magnetic field. We find a significant transfer of atoms from the initial maximally polarized state to the next Zeeman component at magnetic fields of the order of tens of milligauss. We propose an experiment in which such an optimization scheme could  enable the observation of the Einstein-de Haas effect in a rubidium atom condensate.

\end{abstract}

\maketitle


Alkali-metal-atom condensates typically do not behave as dipolar systems as opposed to chromium \cite{52Cr1,52Cr2}, dysprosium \cite{Dy}, or erbium \cite{Er} condensates. This is because the magnetic dipole moment of rubidium atoms is very small. However, there are  some theoretical proposals suggesting that under special conditions dipolar interactions may dominate other interactions present in the condensate, leading to the observable effects \cite{Yi_1,Saito_1,Yi_2,Swislocki_0}. Sometimes, according to the theory, the dipole-dipole interaction can play even crucial role in the condensate dynamics \cite{KG_1}.

The first experimental observation of dipolar interactions in a rubidium condensate was reported in \cite{Vengalattore}. In this experiment the decay of the spin helical structure towards spatially modulated spin domains was observed. Although this process was attributed to the dipolar interactions a theoretical explanation of the role of dipole-dipole forces is still not complete \cite{Kawaguchi,Stamper}.

The recent experiment reported in \cite{Eto} follows the proposal described in Ref. \cite{Saito_1}. According to \cite{Saito_1} the rubidium condensate subject to an external magnetic field (of the order of tens of milligauss) with some field gradient develops the helical spin structure. This helical structure is then modulated by the very weak dipole-dipole interaction (effective magnetic field produced by the magnetic dipoles is about 10 microgauss). Larmor precession of spins around the effective magnetic field together with the movement caused by the field gradient leads to a modulation of the longitudinal magnetization and to a double peak structure in the atomic distributions. Both these effects have been observed in the experiment \cite{Eto}. As numerics shows they are not seen when the magnetic dipolar interaction is turned off.

Even more spectacular, in our opinion, demonstration of dipolar interaction in alkali-metal-atom condensates would be the observation of the Einstein-de Haas effect \cite{EdH}. It is, indeed, remarkable to force the gaseous media into the rotational motion with the help of the magnetic field only. Such a possibility proves the fundamental relation between the spin and the orbital degrees of freedom. In  gaseous media this possibility was first investigated for chromium condensates \cite{Ueda_1,Santos}. For rubidium atoms the effect was first considered in \cite{KG_1}. In gaseous rubidium the Einstein-de Haas effect becomes possible only because there exist resonances which amplify the transfer of atoms between different  Zeeman components \cite{KG_1,Swislocki_1}. It has been recently shown that dipolar resonances occur in chromium condensates \cite{Swislocki_2} as well.

Unfortunately, in alkali-metal-atom condensates dipolar resonances occur at very low magnetic fields, typically of the order of tens of microgauss. Although such ultralow magnetic fields are already accessible at the laboratories \cite{Laburthe_1}, the experiment is nevertheless demanding. It was proposed to replace the static magnetic field by the oscillating one to shift the dipolar resonance towards higher magnetic fields of the order of milligauss \cite{KG_2}. Here, we indicate another route. First, we suggest to use very tight traps of frequencies in the kHz range. Moreover the value of the external magnetic field should vary slowly in time. Because resonant value of the magnetic field depends on the populations of the Zeeman components involved in the process we suggest to change the external magnetic field in time in the way allowing the system to follow the static resonance conditions corresponding to the temporary value of populations. One of the advantages of our method is that the external magnetic field never crosses the zero value.

We do calculations for a rubidium spinor condensate in the $F=1$ hyperfine state within the mean-field approximation. In this approximation the condensate wave function $\psi ({\bf r})=(\psi_1({\bf r}), \psi_0({\bf r}), \psi_{-1}({\bf r}))^T$ fulfills the following equation
\begin{eqnarray}
i\hbar \frac{\partial}{\partial t}\,  \psi ({\bf r}) = ({\cal{H}}_{sp} + {\cal{H}}_c + {\cal{H}}_d )\, \psi ({\bf r})  \,.
\label{eqmot}
\end{eqnarray}
The effective Hamiltonian consists of three terms. The first one, ${\cal{H}}_{sp}=-\frac{\hbar^2}{2 m} \nabla^2 + V_{ext} - \gamma \hbar m_F B$, is the single-particle contribution including the kinetic, potential, and Zeeman energies. In this therm $m$ is a mass of the atom, $V_{ext}$ -- external potential (harmonic trap in our case), $\gamma=-(1/2)\mu_B/\hbar$ is a gyromagnetic coefficient where $\mu_B$ is a Bohr magneton, $m_F=-1,0,1$ are Zeeman sublevels and B is an external magnetic field.  The second term results from the short-range interactions between atoms, whereas the third one is the contribution corresponding to the long-range dipolar interaction. Detailed expressions for ${\cal{H}}_{c}$ and ${\cal{H}}_{d}$ are discussed in the Appendix \ref{first} (also in Ref. \cite{Swislocki_1}). 

For trap frequencies in the kHz region the single particle trap energy is the largest energy scale in the process. The two-body interaction energies, namely the contact and dipole-dipole ones, can be treated as perturbations. The dipole-dipole energy is the smallest energy scale. Dipolar interactions couple the relevant Zeeman components of atoms and lead to a spin dynamics. 

Solving Eq. (\ref{eqmot}) in the case of a time-independent magnetic field reveals that the transfer of atoms between Zeeman components has a resonant character \cite{KG_1,Swislocki_1}. The transfer occurs only within narrow intervals of the values of the external magnetic field. It happens because of the conservation of the total angular momentum and energy. The spin and orbital angular momentum contribute to the total angular momentum and are not independent but are coupled to each other. While going to the other Zeeman state the atoms acquire the orbital angular momentum. Since the orbital motion of atoms is quantized, the particular amount of energy is required for orbital motion. This energy is the Zeeman energy mainly. 

To estimate the Zeeman energy (and hence the resonant magnetic field) allowing for energy conservation and the resonant transfer we can consider the single particle energy only neglecting two-body corrections. In an axial trap the  single particle energies are $(2n_r+|m_{rot}|+1)\hbar\omega_{\perp} + (n_z+1/2)\hbar\omega_z $, where $n_r$, $n_z$, and $m_{rot}$ are radial, axial, and rotational quantum numbers, respectively. The harmonic potential itself is characterized by the radial and axial frequencies: $\omega_{\perp}$ and $\omega_{z}$. In our case the atoms are initially in its orbital ground state therefore up to the leading terms, the resonant values of magnetic fields (or Zeeman shifts) are related to a different final excitation energies in a trap. The resonances can be labeled by quantum numbers $(n_r,|m_{rot}|,n_z)$. Note, that in the transfer between different Zeeman components  the rotational quantum number $|m_{rot}|$ has to increase from zero to compensate for the spin flip of the atom. The  simple estimation of the resonant magnetic field, $B_{res}$ is:
\begin{equation}
g \mu_{B} B_{res}= (2n_r+|m_{rot}|)\hbar\omega_{\perp} + n_z\hbar\omega_z, 
\end{equation}  
where $g=-1/2$ is the Lande factor and $\mu_B$ is the Bohr magneton.

\begin{figure}[thb] \resizebox{3.0in}{2.0in}
{\includegraphics{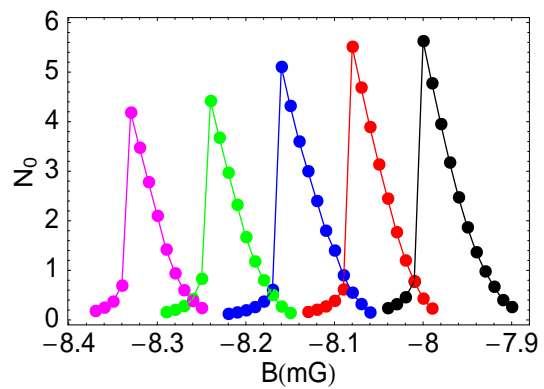}}
\caption{(color online). Maximal transfer, within $50\,$ms, to the $m_F=0$ Zeeman component as a function of the magnetic field. Resonant lines correspond to the initial number of atoms equal to $N_{+1}=100,95,90,85,80$ (from the right to the left). The frequencies of an axially symmetric cigar-shaped trap are $\omega_{x,y}=2\pi \times 6400$Hz and $\omega_z=2\pi \times 1600$Hz. The maximal density in each case is about $10^{15}$cm$^{-3}$. }
\label{res}
\end{figure}

In Fig. \ref{res} we show a number of dipolar resonances for initially polarized, $m_F=1$, rubidium condensate consisted of a small number of atoms strongly confined in an axially symmetric trap. The trap frequencies are of the order of a few kHz. According to the estimation of the resonant energy given in the previous paragraph, the higher trap frequencies lead to  higher resonant magnetic fields. Frequencies in kHz range correspond already to the fields of the order of milligauss. However, because of the three-body losses, the initial number of atoms in $m_F=1$ component is limited to maximally about $100$. This number of atoms corresponds to the density at the center of the trap of about $10^{15}\,$cm$^{-3}$. Each point in Fig. \ref{res} is obtained assuming that the magnetic field is kept constant at a given value and its direction is opposite to the one at which the initial state of a condensate was prepared. Hence, negative values of the magnetic fields are displayed in Fig. \ref{res}.

The question is if for such a small number of atoms the mean field description based on the Gross-Pitaevskii (GP) equation can be used. The answer is positive. In \cite{Sowinski} the exact dynamical solution of two interacting atoms in a harmonic trap was tested against the corresponding solution of the time-dependent GP equation. It was shown that the GP-based description gives correctly the  evolution of the dominant eigenvector of the reduced one particle density matrix. Because weakly interacting condensed systems at low temperatures, as discussed here, are characterized by a single dominant orbital, the GP equations can be safely used.

\begin{figure}[thb] \resizebox{3.0in}{2.0in}
{\includegraphics{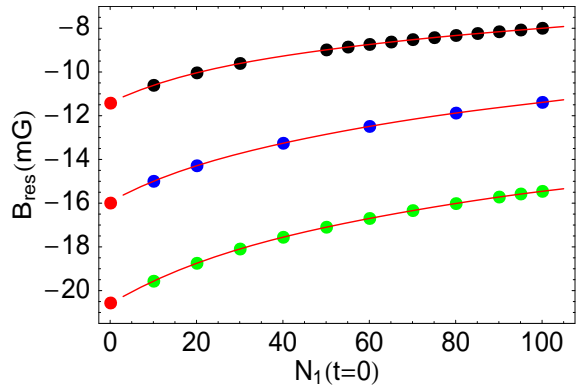}}
\caption{(color online). Resonant magnetic field (in milligauss) as a function of initial number of atoms in $m_F=+1$ Zeeman component. Lines correspond to resonances which differ by an amount of excitation energy deposited in axial direction. The upper, middle, and lower lines (which are fits to numerical data marked by bullets) correspond to resonances responsible for the transfer of atoms to the first, third, and fifth axially excited state, respectively. Red bullets are at the values of resonant magnetic fields assuming no contact interactions are present. Note that the red lines approach the red bullets. }
\label{diagram}
\end{figure}

\begin{figure}[thb] 
\centering
\includegraphics[width=3.6cm]{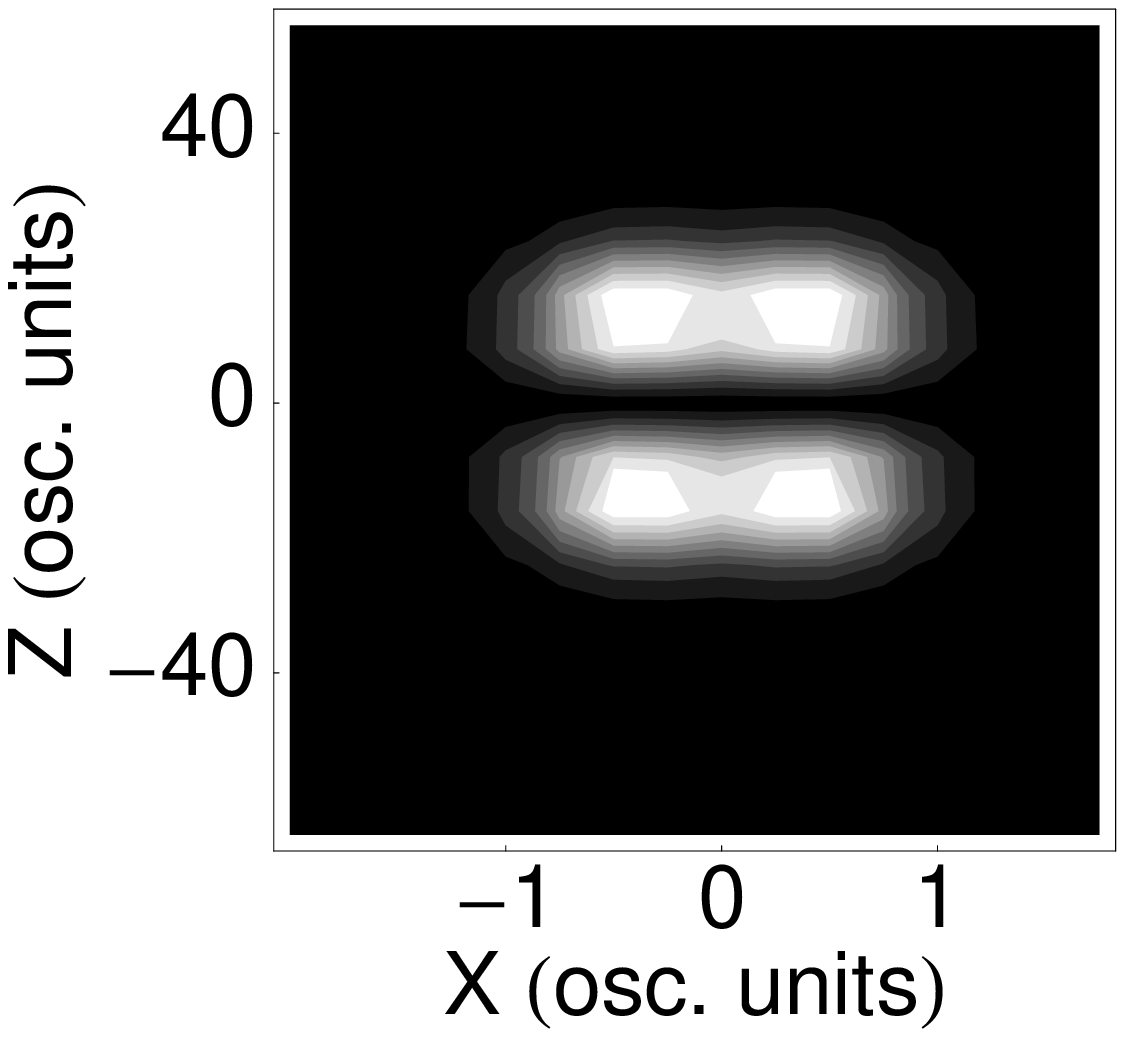} \hfill
\includegraphics[width=3.6cm]{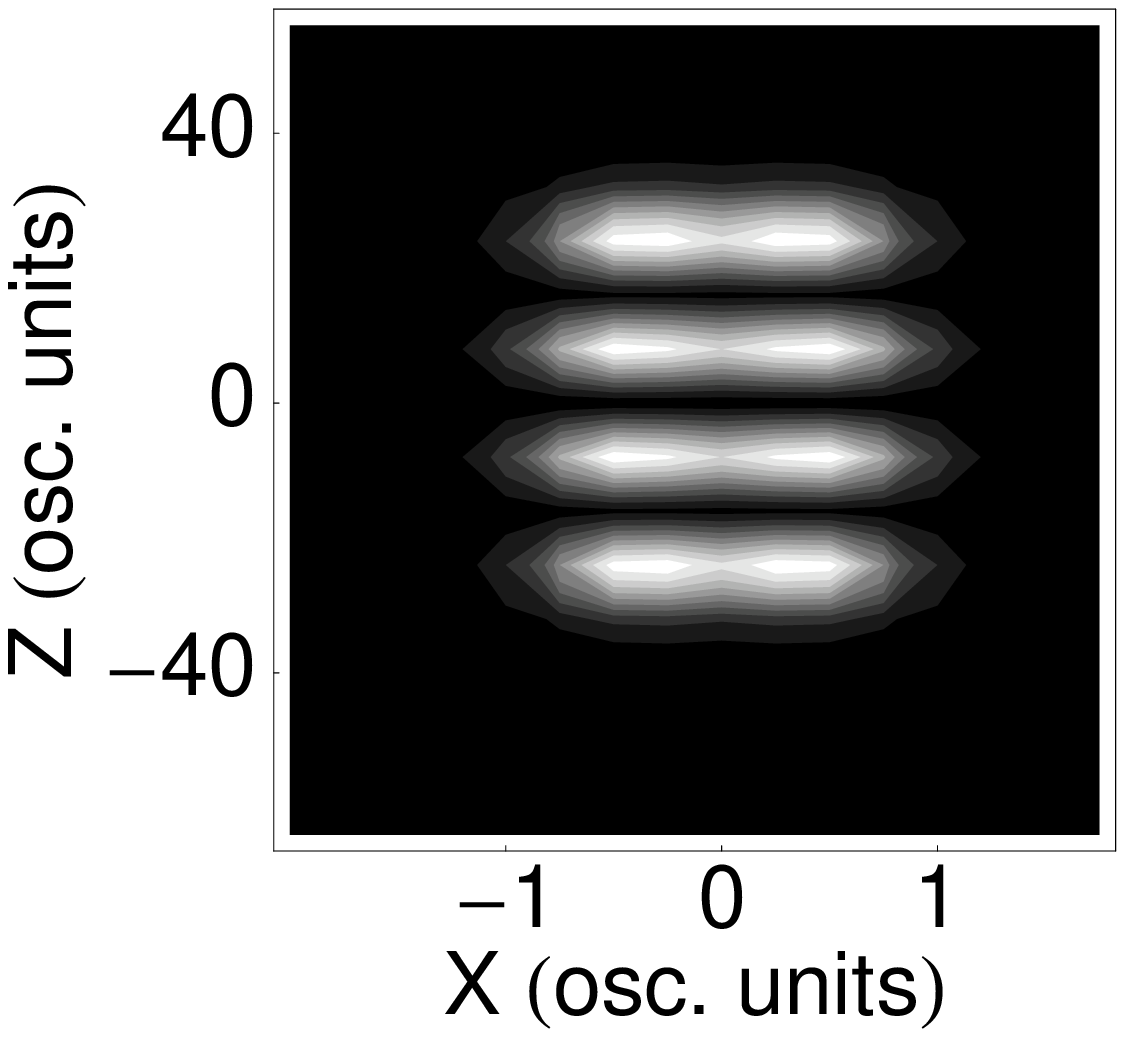}
\caption{Typical density of $m_F=0$ component, integrated along the radial direction, corresponding to the upper (left frame) and middle (right frame) sets of points in Fig. \ref{res}. }
\label{density}
\end{figure}

In Fig. \ref{diagram} we plot the resonant magnetic field as a function of initial number of atoms in $m_F=+1$ Zeeman state. The upper set of points correspond to the resonances showed in Fig. \ref{res}. They are characterized by the following quantum numbers: $m_{rot}=1$, $n_{r}=0$, and $n_{z}=1$. The rotational quantum number equals one which means that a singly-quantized vortex is created in the $m_F=0$ component. Moreover, nonzero value of the axial quantum number tells us that the $m_F=0$ state has some spatial structure along the $z$ direction. Indeed, in the left frame of Fig. \ref{density}, which is the radially integrated density, two rings are visible. The right frame depicts the density pattern typical for the resonance represented by the middle set of points in Fig. \ref{diagram}. The third resonance displayed in Fig. \ref{diagram} (the lowest line) has even more rings in the axial direction. According to the approximate formula for the resonant magnetic field, its value for the upper resonance in Fig. \ref{diagram} is given by $\hbar\omega_{\perp}+\hbar\omega_z\approx -11.4\,$mG and is marked by the red (most left) bullet. Solid lines in Fig. \ref{diagram} approach red (most left) bullets as it should be. It means that the shift in the values of the resonant magnetic field is caused by the contact interaction between atoms.

\begin{figure}[htb] \resizebox{3.0in}{2.0in}
{\includegraphics{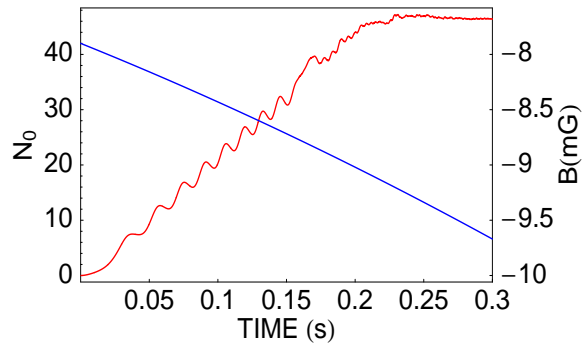}}
\caption{(color online). Population of $m_F=0$ Zeeman component (red line) and the magnetic field (blue line) as functions of time. The initial number of atoms in $m_F=+1$ state is $N_{+1}=100$. The value of the magnetic field changes in time as $B(t)=B_{ini}-\alpha t -\beta t^2$, where $B_{ini}=-7.9$mG, $\alpha=5$mG/s, and $\beta=3$mG$/s^2$.  }
\label{trans}
\end{figure}

\begin{figure}[thb] \resizebox{3.0in}{2.0in}
{\includegraphics{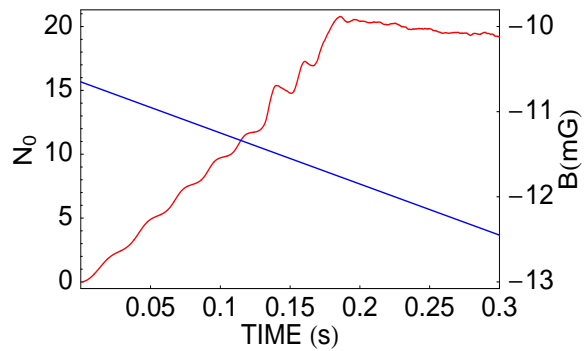}}
\caption{(color online). Population of the $m_F=0$ Zeeman component (red line) and the magnetic field (blue line) as functions of time. The initial number of atoms in $m_F=+1$ state is $N_{+1}=50$. The frequency of a spherically symmetric trap is $\omega_{x,y,z}=2\pi \times 5000$Hz which results in the maximal density of about $10^{15}$cm$^{-3}$. The value of the magnetic field changes in time as $B(t)=B_{ini}-\alpha t$, where $B_{ini}=-10.65$mG and $\alpha=6$mG/s.}
\label{trans1}
\end{figure}

\begin{figure}[thb] \resizebox{3.0in}{2.0in}
{\includegraphics{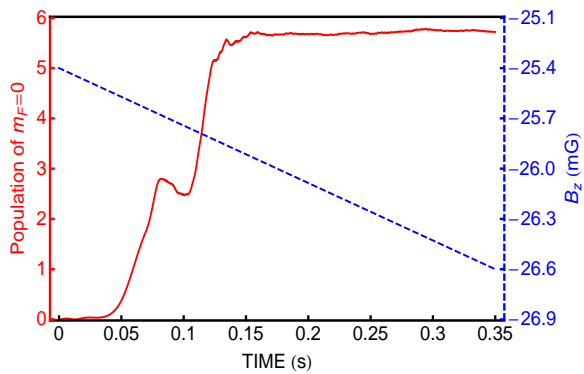}}
\caption{(color online). Population of $m_F=0$ Zeeman component (red solid line) and the magnetic field (blue dashed line) as functions of time. The initial number of atoms in $m_F=+1$ state is $N_{+1}=50$. The frequency of a spherically symmetric trap is $\omega_{x,y,z}=2\pi \times 5000$Hz. The value of the magnetic field changes in time as $B(t)=B_{ini}-\alpha t$, where $B_{ini}=-25.4$mG and $\alpha=1.2$mG/s.  }
\label{trans2}
\end{figure}

\begin{figure}[thb] 
\centering
\includegraphics[width=3.6cm]{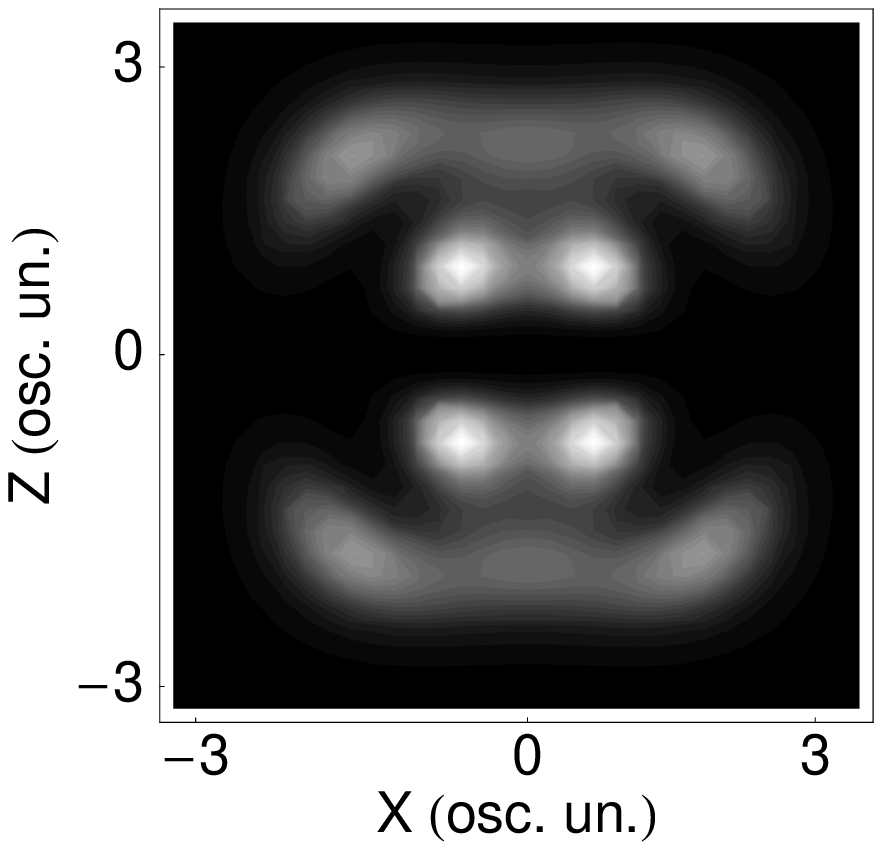} \hfill
\includegraphics[width=3.6cm]{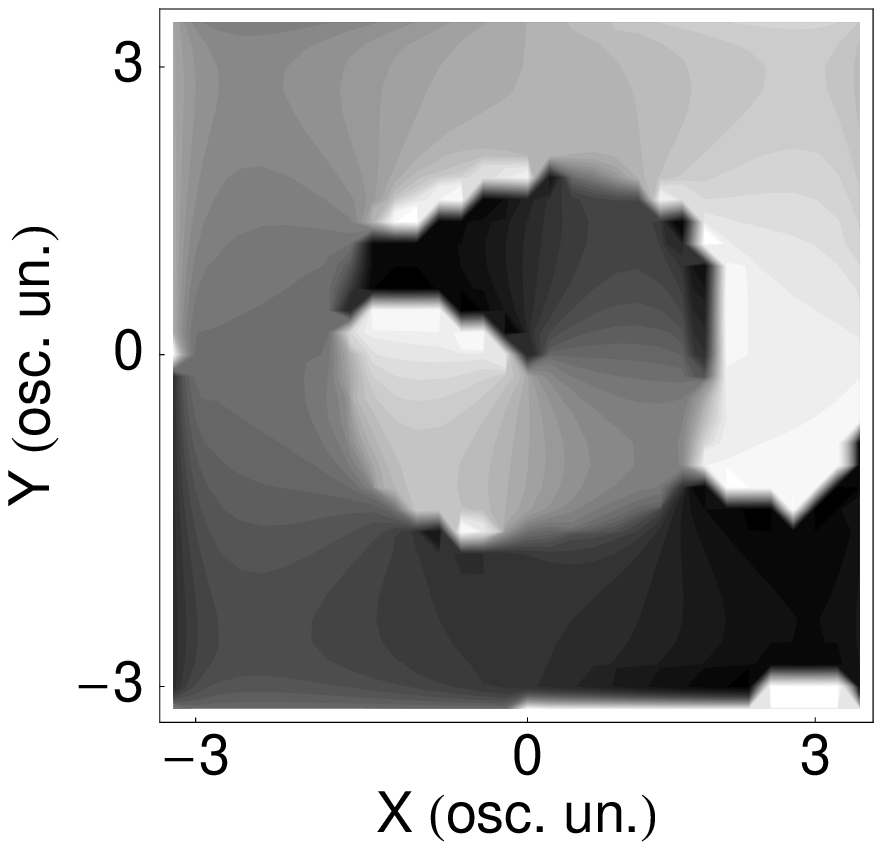}
\caption{Typical density in the $xz$ plane integrated along the $y$ direction (left frame) and the phase in the $xy$ plane (right frame) of the $m_F=0$ component, corresponding to the resonance, as in Fig. \ref{trans2}, characterized by the quantum numbers $n_r=m_{rot}=n_z=1$. }
\label{density2}
\end{figure}

Now, the question is can we change in time the value of the magnetic field in such a way that the system follows any of the solid lines plotted in Fig. \ref{diagram}. Indeed, it is possible as it is demonstrated in Fig. \ref{trans}. Here we show the population of the $m_F=0$ component while the magnetic field varies in time as $B(t)=B_{ini}-\alpha t -\beta t^2$, where $B_{ini}=-7.9$mG, $\alpha=5$mG/s, and $\beta=3$mG$/s^2$. After about $200\,$ms almost half of the atoms is transferred from $m_F=1$ to the $m_F=0$ Zeeman state. Further improvement in the transfer is possible but requires detailed analysis of the position of the dipolar resonance in the case when the nearest spin state is already significantly populated. From the point of view of the observation of the Einstein-de Haas effect it is even better to work with the spherically symmetric traps. Fig. \ref{trans1} shows that also in this case a large transfer (again almost $50\%$) is possible at the magnetic field of dozen milligauss. For spherically symmetric traps, similarly  like for cigar-shaped traps, other kinds of resonances can be also excited. Fig. \ref{trans2} proves that a significant transfer of atoms (here, more than $10\%$) corresponding to the resonance which is characterized by the radial excitations is also manageable (Fig. \ref{density2} shows the density and phase pattern for this case).

Based on the findings discussed so far we propose the following experiment which allows the observation of the Einstein-de Haas effect in a rubidium condensate. We have seen that for tight confinement a significant transfer of atoms from $m_F=+1$ to $m_F=0$ component is possible in the magnetic field of the order of tens milligauss. For that the value of the magnetic field needs to be changed in time to tune the system to the temporal position of the dipolar resonance corresponding to the transient population of $m_F=+1$ state. The transfer to $m_F=0$ Zeeman state is high enough to allow the separation of various Zeeman components by using the Stern-Gerlach technique. Also the magnetic field does not cross the zero value as opposed to the methods applying the oscillating magnetic fields. To amplify the signal instead of a single trap a linear chain of several sites each confining tens of rubidium atoms should be considered. Applying the magnetic field perpendicular or parallel to the axis of a chain of microtraps and doing, after separation of components, absorption imaging along this axis one should be able to observe density patterns proving the realization of the Einstein-de Haas effect.

In conclusion, we have studied the dynamics of a rubidium spinor condensate in a time-dependent magnetic field. Optimizing the change of the value of the magnetic field we are able to follow the resonance despite of the atomic losses due to dipolar interaction. Therefore, the significant number of atoms is transferred from the initial to the final Zeeman state. Since the atoms are tightly confined the value of the resonant magnetic field can be pushed in the range of tens of milligauss. We propose an experiment in which the Einstein-de Haas effect as a spectacular demonstration of dipolar interaction in alkali-metal-atoms condensates could be observed.

\acknowledgments

This work was supported by the National Science Center grants 
No. DEC-2011/01/B/ST2/05125, DEC-2011/03/D/ST2/01938 (T.\'S.),  No. DEC-2011/01/D/ST2/02019 (M.B., M.G.).

\appendix
\section{Equation of motion}
\label{first}

In the second quantization the system we study is described by the following Hamiltonian:

\begin{eqnarray}
&&H = \int d^3r \left[ \hat{\psi}^{\dagger}_i(\mathbf{r}) \left(-\frac{\hbar^2}{
2 m} 
\nabla^2 + V_{ext}(\mathbf{r}) \right)   \hat{\psi}_i(\mathbf{r})    
\right.  \nonumber  \\
&&\left. -\gamma \hat{\psi}^{\dagger}_i(\mathbf{r})\, \mathbf{B F}_{ij}\, 
\hat{\psi}_j(\mathbf{r}) + \frac{c_0}{2}\, \hat{\psi}^{\dagger}_j(\mathbf{r})
\hat{\psi}^{\dagger}_i(\mathbf{r})
\hat{\psi}_i(\mathbf{r}) \hat{\psi}_j(\mathbf{r}) \right.   \nonumber  \\
&&\left. +\frac{c_2}{2}\, \hat{\psi}^{\dagger}_k(\mathbf{r})
\hat{\psi}^{\dagger}_i(\mathbf{r})\,
\mathbf{F}_{ij} \mathbf{F}_{kl}\, \hat{\psi}_j(\mathbf{r}) \hat{\psi}_l(\mathbf{
r})
\right]   \nonumber  \\
&&+ \frac{1}{2}\int d^3r\, d^3r' \hat{\psi}^{\dagger}_k(\mathbf{r})
\hat{\psi}^{\dagger}_i(\mathbf{r}') V^d_{ij,kl}(\mathbf{r}-\mathbf{r}')
\hat{\psi}_j(\mathbf{r}') \hat{\psi}_l(\mathbf{r})   \,,  \nonumber  \\
\label{Ham}
\end{eqnarray}
where repeated indices (each of them going through the values $+1$, $0$, and $−1$) are to be summed over. To derive Eq. (\ref{eqmot}) one needs to calculate the commutators $[\hat{\psi}_i(\mathbf{r}),H]$, where $H$ is given by (\ref{Ham}). The first line in (\ref{Ham}) describes the kinetic and the trapping energies. The second term describes the interaction with the external magnetic field {\bf B} with $\gamma$ being the gyromagnetic coefficient which relates the effective magnetic moment with the hyperfine angular momentum.

The term ${\cal{H}}_c$ in Eq. (\ref{eqmot}), which is related to the contact interactions has the following diagonal elements
\begin{eqnarray}
{\cal{H}}_{c 11} &=& (c_0+c_2)\, \hat{\psi}^{\dagger}_1 \hat{\psi}_1 
+(c_0+c_2)\, \hat{\psi}^{\dagger}_0 \hat{\psi}_0   \nonumber \\
&+&  (c_0-c_2)\, \hat{\psi}^{\dagger}_{-1} \hat{\psi}_{-1}  \nonumber  \\
{\cal{H}}_{c 00} &=&(c_0+c_2)\, \hat{\psi}^{\dagger}_1\hat{\psi_1} +
c_0\, \hat{\psi}^{\dagger}_0\hat{\psi_0}  \nonumber \\
&+& (c_0+c_2)\, \hat{\psi}^{\dagger}_{-1}\hat{\psi_{-1}} \nonumber \\
{\cal{H}}_{c -1-1} &=&(c_0-c_2)\, \hat{\psi}^{\dagger}_1\hat{\psi}_1
+(c_0+c_2)\, \hat{\psi}^{\dagger}_0\hat{\psi}_0  \nonumber \\
&+& (c_0+c_2)\, \hat{\psi}^{\dagger}_{-1} \hat{\psi}_{-1}  \;.
\label{Hc}
\end{eqnarray}
These elements describe collisions of atoms that preserve the projection of the spin of each atom. The off-diagonal elements, on the other hand, are responsible for collisions changing the atomic spin projections but conserving the projection of the total spin. They are equal
\begin{eqnarray}
{\cal{H}}_{c 10} &=& c_2\hat{\psi}^{\dagger}_{-1} \hat{\psi}_0  \nonumber \\
{\cal{H}}_{c 0-1} &=& c_2\hat{\psi}^{\dagger}_0\hat{\psi}_1  \nonumber \\
{\cal{H}}_{c 1-1} &=& 0  \;.
\end{eqnarray}

The terms with coefficients $c_0$ and $c_2$ can be expressed with the help of scattering lengths $a_0$ and $a_2$, where $c_0=4\pi\hbar^2(a_0+a_2)/3m$ and $c_2=4\pi\hbar^2(a_2-a_0)/3m$ \cite{scatt}. The scattering lengths $a_0$ and $a_2$ determine the collisions of atoms in a channel of the total spin 0 and 2, respectively. According to \cite{scatt2} the $a_0=5.387$nm and $a_2=5.313$nm. The $\bf F$ are spin-1 matrices (Eq. \ref{F}):

\begin{displaymath}
F_x=\frac{\hbar}{\sqrt{2}}\left(\begin{array}{ccc} 0 & 1 & 0\\
1 & 0 & 1 \\ 0 & 1 & 0 \end{array} \right), \,\, F_y=\frac{\hbar}{\sqrt{2}}\left(\begin{array}{ccc} 0 & -i & 0\\
i & 0 & -i \\ 0 & i & 0 \end{array} \right),
\end{displaymath}
\begin{eqnarray}
F_z=\hbar \left(\begin{array}{ccc} 1 & 0 & 0\\
0 & 0 & 0 \\ 0 & 0 & -1 \end{array} \right).
\label{F}
\end{eqnarray}

The dipolar interactions ${\cal{H}}_d$ in Eq. (\ref{eqmot}) is as follows:

\begin{eqnarray}
{\cal{H}}_{d ij}(\mathbf{r}) &=& \int d^3r' \hat{\psi}_k^{\dagger}(\mathbf{r}')
V^d_{ij,kl}(\mathbf{r}-\mathbf{r}')   \hat{\psi}_l(\mathbf{r}')   \;,
\label{Hd}
\end{eqnarray}
where
\begin{eqnarray}
&&V^d_{ij,kl}(\mathbf{r}-\mathbf{r}') = \frac{\gamma^2}{|\mathbf{r}-\mathbf{r}'|^3}
\mathbf{F}_{ij} \mathbf{F}_{kl}  \nonumber \\
&&- \frac{3\gamma^2}{|\mathbf{r}-\mathbf{r}'|^5}
[\mathbf{F}_{ij}  (\mathbf{r}-\mathbf{r}')] [\mathbf{F}_{kl}  (\mathbf{r}-\mathbf{r}')]  
\,\,.
\label{Vd}
\end{eqnarray}
There are two elements of ${\cal{H}}_d$ matrix that are independent:
\begin{eqnarray}
&&{\cal{H}}_{d 11}(\mathbf{r}) = \hbar^2 \gamma^2 \int d^3r' \left[
\frac{1}{|\mathbf{r}-\mathbf{r}'|^3}-3\frac{(z-z')^2}{|\mathbf{r}-\mathbf{r}'|^5}
\right] \nonumber \\
&&\times (\hat{\psi}_1^{\dagger} \hat{\psi}_1-\hat{\psi}_{-1}^{\dagger} \hat{\psi}_{-1})   
\nonumber \\
&& -3 \frac{\hbar^2 \gamma^2}{\sqrt{2}}
\int d^3r' \frac{z-z'}{|\mathbf{r}-\mathbf{r}'|^5}[(x-x') - i (y-y')]  \nonumber \\
&&\times (\hat{\psi}_1^{\dagger}\hat{\psi}_0+\hat{\psi}_0^{\dagger}\hat{\psi}_{-1})    \nonumber \\
&& -3 \frac{\hbar^2 \gamma^2}{\sqrt{2}}
\int d^3r' \frac{z-z'}{|\mathbf{r}-\mathbf{r}'|^5}[(x-x') + i (y-y')] \nonumber \\
&&\times (\hat{\psi}_0^{\dagger}\hat{\psi}_1+\hat{\psi}_{-1}^{\dagger}\hat{\psi}_0)
\label{Hd_11}
\end{eqnarray}
and
\begin{eqnarray}
&&{\cal{H}}_{d 10}(\mathbf{r}) = -3 \frac{\hbar^2 \gamma^2}{\sqrt{2}} \int d^3r' 
\frac{[(x-x')-i(y-y')](z-z')}{|\mathbf{r}-\mathbf{r}'|^5} \nonumber \\
&& \times (\hat{\psi}_1^{\dagger} \hat{\psi}_1-\hat{\psi}_{-1}^{\dagger} \hat{\psi}_{-1})    
\nonumber \\
&& -\frac{3}{2} \hbar^2 \gamma^2 \int d^3r'
\frac{[(x-x')-i(y-y')]^2}{|\mathbf{r}-\mathbf{r}'|^5}
(\hat{\psi}_1^{\dagger}\hat{\psi_0}+\hat{\psi_0}^{\dagger}\hat{\psi}_{-1})   \nonumber \\
&& +\hbar^2 \gamma^2 \int d^3r' \left[
\frac{1}{|\mathbf{r}-\mathbf{r}'|^3}-\frac{3}{2}\frac{(x-x')^2+(y-y')^2}
{|\mathbf{r}-\mathbf{r}'|^5} \right]   \nonumber \\
&& \times (\hat{\psi}_0^{\dagger}\hat{\psi}_1+\hat{\psi}_{-1}^{\dagger}\hat{\psi}_0)   \;.
\label{Hd_10}
\end{eqnarray}
Moreover,
\begin{eqnarray}
&&{\cal{H}}_{d 0-1} = {\cal{H}}_{d 10}  , \;\;\;
{\cal{H}}_{d -1-1} = - {\cal{H}}_{d 11}   \nonumber \\
&& {\cal{H}}_{d 1-1} = {\cal{H}}_{d 00} = 0  \,.
\label{Hd_con}
\end{eqnarray}

All the ${\cal{H}}_{dij}$ terms are responsible for the change of the total spin projection of colliding atoms.

Finally, assuming the macroscopic occupation of all spinor components, we replace the field operators by complex functions. This corresponds to the mean-field approximation.

\end{document}